\documentstyle[12pt]{article}                    
\newfont{\ffont}{msym10}                          
\newcommand{\beq}{\begin{equation}}               
\newcommand{\eeq}{\end{equation}}                 
\newcommand{\bqry}{\begin{eqnarray}}              
\newcommand{\eqry}{\end{eqnarray}}                
\newcommand{\bqryn}{\begin{eqnarray*}}            
\newcommand{\eqryn}{\end{eqnarray*}}              
\newcommand{\preprint}[1]{\begin{table}[t]        
            \begin{flushright}                    
            \begin{large}{#1}\end{large}          
            \end{flushright}                      
            \end{table}}                          
\newcommand{\PD}[2]                               
    {\frac{\partial^{#2}}{\partial #1^{#2}}}      
\begin{document}
\preprint{LA-UR-97-793}
\title{On $D$-Wave Meson Spectroscopy \\ and the $K^\ast (1410)$-$K^\ast (
1680)$ Problem}
\author{\\ L. Burakovsky\thanks{E-mail: BURAKOV@PION.LANL.GOV} \
and \ T. Goldman\thanks{E-mail: GOLDMAN@T5.LANL.GOV} \
\\  \\  Theoretical Division, MS B285 \\  Los Alamos National Laboratory \\ 
Los Alamos, NM 87545, USA \\}
\date{ }
\maketitle
\begin{abstract}
The mass spectrum of $D$-wave mesons is considered in a nonrelativistic 
constituent quark model. The results show a common mass degeneracy of the
isovector and isodoublet states of the 1 $^3D_1$ and 1 $^3D_3$ nonets, and 
suggest therefore that the $K^\ast (1680)$ cannot be the $I=1/2$ member of the
1 $^3D_1$ nonet. They also suggest that the $\eta _2(1870),$ presently omitted
from the Meson Summary Table, should be interpreted as the $I=0$ $s\bar{s}$ 
state of the 1 $^1D_2$ nonet.
\end{abstract}

\bigskip
\section{Introduction}
The existence of a gluon self-coupling in QCD suggests that, in addition to 
the conventional $q\bar{q}$ states, there may be non-$q\bar{q}$ mesons: bound 
states including gluons (gluonia and glueballs, and $q\bar{q}g$ hybrids) and 
multiquark states \cite{1}. Since the theoretical guidance on the properties 
of unusual states is often contradictory, models that agree in the $q\bar{q}$
sector differ in their predictions about new states. Among the naively 
expected signatures for gluonium are \hfil\break
i) no place in $q\bar{q}$ nonet, \hfil\break
ii) flavor-singlet coupling, \hfil\break
iii) enhanced production in gluon-rich channels such as $J/\Psi (1S)$ decay, 
\hfil\break iv) reduced $\gamma \gamma $ coupling, \hfil\break v) exotic 
quantum numbers not allowed for $q\bar{q}$ (in some cases). \hfil\break
Points iii) and iv) can be summarized by the Chanowitz $S$ parameter \cite{Cha}
$$S=\frac{\Gamma (J/\Psi (1S)\rightarrow \gamma X)}{{\rm PS} (J/\Psi (1S)
\rightarrow \gamma X)}\times \frac{{\rm PS} (X\rightarrow \gamma \gamma )}{
\Gamma (X\rightarrow \gamma \gamma )},$$
where PS stands for phase space. $S$ is expected to be larger for gluonium 
than for $q\bar{q}$ states. Of course, mixing effects and other dynamical 
effects such as form-factors can obscure these simple signatures. Even if the 
mixing is large, however, simply counting the number of 
observed states remains a clear signal for non-exotic non-$q\bar{q}$ states.
Exotic quantum number states $(0^{--},0^{+-},1^{-+},2^{+-},\ldots )$ would be
the best signatures for non-$q\bar{q}$ states. It should be also emphasized 
that no state has yet unambiguously been identified as gluonium, or as a 
multiquark state, or as a hybrid. 

In this paper we shall discuss $D$-wave meson states, the interpretation of 
which as members of conventional quark model $q\bar{q}$ nonets encounters 
difficulties \cite{enigmas}. We shall be concerned with the four meson nonets 
which have the following $q\bar{q}$ quark model assignments, according to the 
most recent Review of Particle Physics \cite{pdg}:\hfil\break
1) $\;1\; ^1D_2$ $J^{PC}=2^{-+},$ $\;\pi _2(1670),\;\;\eta _2^{'}($ ? $),\;\;
\eta _2($ ? $),\;\;K_2(1770)$\hfil\break
2) $\;1\; ^3D_1$ $J^{PC}=1^{--},$ $\;\rho (1700),\;\;\omega (1600),\;\;\phi ($ ? $),\;\;\;K^\ast (1680)$\hfil\break
3) $\;1\; ^3D_2$ $J^{PC}=2^{--},$ $\;\rho _2($ ? $),\;\;\;\omega _2($ ? $),\;\;
\;\phi _2($ ? $),\;\;K_2^{'}(1820)$\hfil\break
4) $\;1\; ^3D_3$ $J^{PC}=3^{--},$ $\;\rho _3(1690),\omega _3(1670),\phi _3(
1850),K_3^\ast (1780),$\hfil\break
and start with a discussion of the corresponding two problems associated with
the isodoublet channel of these nonets. One of them is related to the $K^\ast (
1410)-K^\ast (1680)$ problem, the other to possible $^1D_2-^3D_2$ mixing in
the $I=1/2$ channel.
 \\   \\ 
The two mesons, $K^\ast (1680)$ (with mass $1714\pm 20$ MeV and width $323\pm
110$ MeV) and $K^\ast (1410)$ $(1412\pm 12$ MeV, $227\pm 22$ MeV) are currently
assigned to the 1 $^3D_1$ and 2 $^3S_1$ nonets, respectively (the latter, 2 $^
3S_1$ $J^{PC}=1^{--},$ $\rho (1450),$ $\omega (1420),$ $\phi (1680),$ $K^\ast 
(1410),$ has the same flavor quantum numbers as the former), although, as the 
Particle Data Group (PDG) states, ``the $K^\ast (1410)$ could be replaced by 
the $K^\ast (1680)$ as the 2 $^3S_1$ state'' \cite{pdg1}. The problem with 
these mesons is that the $K^\ast (1410)$ seems too light to be the 2 $^3S_1$ 
state, even if one takes into account possible $2\;^3S_1-1\;^3D_1$ mixing. 
Similarly, the $K^\ast (1680)$ seems too light to be the 1 $^3D_1.$ One may  
doubt even the existence of the $K^\ast (1410),$ as suggested first by 
T\"{o}rnqvist \cite{To}, since it (as well as the $K^\ast (1680))$ has been
observed by only one group, LASS \cite{LASS}, although with superior 
statistics, in partial wave analyses under the much stronger $K_2^\ast (1430)$
and $K_0^\ast (1430).$ Two older experiments \cite{Etkin,older} quote a 
considerably higher mass, $\simeq 1500$ MeV. In addition, its $K\pi $ 
branching ratio is suspiciously small, only $(6.6\pm 1.3)$\%. On the other 
hand, the $K^\ast (1680)$ has a suspiciously large total width $(\sim 400)$ 
MeV, much larger than typical hadron widths, and a natural suspicion would be 
that it is really composed of two states of normal width $(\sim 150-200$ MeV)
\cite{To}, quite analogously to what has been suggested to be the case for the
$\rho (1600)$ and $\omega (1600)$ which have been resolved into $\rho (1450)$
plus $\rho (1700)$ and $\omega (1420)$ plus $\omega (1600)$ \cite{split}. The 
masses of the two states contained in the $K^\ast (1680)$ were determined in 
ref. \cite{To}to be 2 $^3S_1(\approx \!1608)$ and 1 $^3D_1(\approx \!
1784),$ from the requirement that the both fit the corresponding Regge 
trajectories. This is in agreement with the values obtained by Godfrey and 
Isgur in a relativized quark model \cite{GI}, 2 $^3S_1(1580),$ 1 $^3D_1(1780).$
An older experiment on the $K^\ast (1680)$ quotes a mass of the same order,
$\sim 1800$ MeV \cite{Etkin}. 
 \\   \\
Theoretically, for the four $(n,L)$-wave meson nonets, the isoscalar and 
isovector members of the $n\;^3L_L$ and $n\;^1L_L$ nonets with the same 
charge cannot mix, since they have opposite $C$- and $G$-parity, as long as 
one neglects $SU(2)_I$ breaking. However, their isodoublet counterparts 
(strange, charmed, ... mesons) do not possess definite $C$-parity and, 
therefore, can in principle mix when only $SU(3)$ flavor symmetry is broken. 
This type of mixing can take place for all $L\geq 1$ mesons, as follows,
\beq
\left( \begin{array}{c}
Q_{high} \\ 
Q_{low}
\end{array} \right) =\left( \begin{array}{cc}
\cos \theta _{nL} & \sin \theta _{nL} \\
-\sin \theta _{nL} & \cos \theta _{nL}
\end{array} \right) \left( \begin{array}{cc}
n\;^1L_L \\
n\;^3L_L
\end{array} \right) ,
\eeq
where $Q$ stands for the $K,D,D_s,$ ... . It is known that this mixing actually
takes place for the $P$-wave mesons where the $I=1/2$ $K_{1A}$ and $K_{
1B}$ states of the 1 $^3P_1$ and 1 $^1P_1$ nonets, respectively, mix, leading
to the physical $K(1270)$ and $K(1400)$ states \cite{K1,Lip}. If such a mixing
is also the case for the $D$-wave mesons, a question suggests itself regarding 
the physical masses of the $I=1/2$ states of the $^3D_2$ and $^1D_2$ nonets, 
which we call $K_{2A}$ and $K_{2B},$ respectively, in the following.
 \\

If the assumption of T\"{o}rnqvist about the $K^\ast (1680)$ \cite{To} is 
correct, one would have simultaneous mass near-degeneracy of the 1 $^3D_1$ and
1 $^3D_3$ meson nonets in the isovector and isodoublet channels, since in this 
case $M(\rho (1700))\approx M(\rho _3(1690)),$ $M(K^\ast (1780))\approx M(K_3^
\ast (1780)).$ As shown in our previous paper \cite{prev}, similar degeneracy 
of the 1 $^3P_0$ and 1 $^3P_2$ nonets is an intrinsic property of $P$-wave 
meson spectroscopy and may be straightforwardly understood in a nonrelativistic
constituent quark model. We now wish to apply this model to the $D$-wave mesons
in order to show that near-degeneracy of the $^3D_3$ and $^3D_1$ nonets 
mentioned above also takes place. We note that this result is a direct 
consequence of the nonrelativistic constituent quark model which we discuss 
below; this mass near-degeneracy of the two nonets does not depend on the 
values of the input parameters, and cannot be considered as a numerical 
coincidence, as the results of, e.g., Godfrey and Isgur \cite{GI}, may be 
viewed (their model finds the values $M(K^\ast )=1780$ MeV, $M(K_3^\ast )=1790$
MeV for the $I=1/2$ 1 $^3D_1$ and 1 $^3D_3$ meson masses). We also expect our
model to provide relevant information on possible $K_{2A}-K_{2B}$ mixing.

\section{Nonrelativistic constituent quark model}
In the constituent quark model, conventional mesons are bound states of a spin
1/2 quark and spin 1/2 antiquark bound by a phenomenological potential which 
has some basis in QCD \cite{LSG}. The quark and antiquark spins combine to 
give a total spin 0 or 1 which is coupled to the orbital angular momentum $L.$
This leads to meson parity and charge conjugation given by $P=(-1)^{L+1}$ and 
$C=(-1)^{L+S},$ respectively. One typically assumes that the $q\bar{q}$ wave 
function is a solution of a nonrelativistic Schr\"{o}dinger equation with the 
generalized Breit-Fermi Hamiltonian\footnote{The most widely used potential 
models are the relativized model of Godfrey and Isgur \cite{GI} for the 
$q\bar{q}$ mesons, and Capstick and Isgur \cite{CI} for the $qqq$ baryons. 
These models differ from the nonrelativistic quark potential model only in
relatively minor ways, such as the use of $H_{kin}=\sqrt{m_1^2+{\bf p}_1^2}+
\sqrt{m_2^2+{\bf p}_2^2}$ in place of that given in (2), the retention of the 
$m/E$ factors in the matrix elements, and the introduction of coordinate
smearing in the singular terms such as $\delta ({\bf r}).$}, $H_{BF},$
\beq
H_{BF}\;\psi _n({\bf r})\equiv \left( H_{kin}+V({\bf p},{\bf r})\right) \psi _
n({\bf r})=E_n\psi _n({\bf r}),
\eeq
where $H_{kin}=m_1+m_2+{\bf p}^2/2\mu -(1/m_1^3+1/m_2^3){\bf p}^4/8,$ $\mu =m_
1m_2/(m_1+m_2),$ $m_1$ and $m_2$ are the constituent quark masses, and to
first order in $(v/c)^2={\bf p}^2c^2/E^2\simeq {\bf p}^2/m^2c^2,$ $V({\bf p},
{\bf r})$ reduces to the standard nonrelativistic result, 
\beq
V({\bf p},{\bf r})\simeq V(r)+V_{SS}+V_{LS}+V_T,
\eeq
with $V(r)=V_V(r)+V_S(r)$ being the confining potential which consists of a
vector and a scalar contribution, and $V_{SS},V_{LS}$ and $V_T$ the spin-spin,
spin-orbit and tensor terms, respectively, given by \cite{LSG}
\beq
V_{SS}=\frac{2}{3m_1m_2}\;{\bf s}_1\cdot {\bf s}_2\;\triangle V_V(r),
\eeq
$$V_{LS}=\frac{1}{4m_1^2m_2^2}\frac{1}{r}\left( \left\{ [(m_1+m_2)^2+2m_1m_2]\;
{\bf L}\cdot {\bf S}_{+}+(m_2^2-m_1^2)\;{\bf L}\cdot {\bf S}_{-}\right\}
\frac{dV_V(r)}{dr}\right. $$
\beq
\left. -\;[(m_1^2+m_2^2)\;{\bf L}\cdot {\bf S}_{+}+(m_2^2-m_1^2)\;{\bf L}\cdot
{\bf S}_{-}]\;\frac{dV_S(r)}{dr}\right) ,
\eeq
\beq
V_T=\frac{1}{12m_1m_2}\left( \frac{1}{r}\frac{dV_V(r)}{dr}-\frac{d^2V_V(r)}{
dr^2}\right) S_{12}.
\eeq
Here ${\bf S}_{+}\equiv {\bf s}_1+{\bf s}_2,$ ${\bf S}_{-}\equiv {\bf s}_1-
{\bf s}_2,$ and
\beq
S_{12}\equiv 3\left( \frac{({\bf s}_1\cdot {\bf r})({\bf s}_2\cdot {\bf r})}{
r^2}-\frac{1}{3}{\bf s}_1\cdot {\bf s}_2\right).
\eeq
For constituents with spin $s_1=s_2=1/2,$ $S_{12}$ may be rewritten in the form
\beq
S_{12}=2\left( 3\frac{({\bf S}\cdot {\bf r})^2}{r^2}-{\bf S}^2\right),\;\;\;
{\bf S}={\bf S}_{+}\equiv {\bf s}_1+{\bf s}_2.
\eeq
Since $(m_1+m_2)^2+2m_1m_2=6m_1m_2+(m_2-m_1)^2,$ $m_1^2+m_2^2=2m_1m_2+(m_2-m_
1)^2,$ the expression for $V_{LS},$ Eq. (5), may be rewritten as follows,
$$V_{LS}=\frac{1}{2m_1m_2}\frac{1}{r}\left[ \left( 3\frac{dV_V(r)}{dr}-
\frac{dV_S(r)}{dr}\right) + \frac{(m_2-m_1)^2}{2m_1m_2}\left(\frac{dV_V(r)}{d
r}-\frac{dV_S(r)}{dr}\right) \right] {\bf L}\cdot {\bf S}_{+}$$
\beq
+\frac{m_2^2-m_1^2}{4m_1^2m_2^2}\;\frac{1}{r}\left( \frac{dV_V(r)}{dr}-\frac{
dV_S(r)}{dr}\right) {\bf L}\cdot {\bf S}_{-}\equiv V_{LS}^{+}+V_{LS}^{-}.
\eeq
Since two terms corresponding to the derivatives of the potentials with respect
to $r$ are of the same order of magnitude, the above expression for 
$V_{LS}^{+}$ may be rewritten as
\beq
V_{LS}^{+}=\frac{1}{2m_1m_2}\frac{1}{r}\left( 3\frac{dV_V(r)}{dr}-\frac{dV_
S(r)}{dr}\right) {\bf L}\cdot {\bf S}\left[ 1+\frac{(m_2-m_1)^2}{2m_1m_2}\;O(
1)\right] .
\eeq

\section{$D$-wave spectroscopy}
We now wish to apply the Breit-Fermi Hamiltonian to the $D$-wave mesons. By
calculating the expectation values of different terms of the Hamiltonian 
defined in Eqs. (4),(8),(9), taking into account the corresponding matrix 
elements $\langle {\bf s}_1\cdot {\bf s}_2\rangle ,$ $\langle {\bf L}\cdot 
{\bf S}\rangle $ and $S_{12}$ \cite{LSG}, one obtains relations similar to
those for the $P$-wave mesons \cite{prev,BGP},
\bqryn
M(^3D_1) & = & M_0+\frac{1}{4}\langle V_{SS}\rangle -3\langle V_{LS}^{+}
\rangle -\frac{1}{2}\langle V_T\rangle , \\
M(^3D_3) & = & M_0+\frac{1}{4}\langle V_{SS}\rangle +2\langle V_{LS}^{+}\rangle
-\frac{1}{7}\langle V_T\rangle , \\  
M(\rho _2) & = & M_0+\frac{1}{4}\langle V_{SS}\rangle -\langle V_{LS}^{+}
\rangle +\frac{1}{2}\langle V_T\rangle , \\
M(\pi _2) & = & M_0-\frac{3}{4}\langle V_{SS}\rangle ,
\eqryn 
$$\left( \begin{array}{c}
M(K_2^{'}) \\ M(K_2) \end{array} \right) =\left( \begin{array}{cc}
M_0+\frac{1}{4}\langle V_{SS}\rangle -\langle V_{LS}^{+}\rangle +\frac{1}{2}
\langle V_T\rangle  & \sqrt{2}\langle V_{LS}^{-}\rangle \\ 
\sqrt{2}\langle V_{LS}^{-}\rangle  & M_0-\frac{3}{4}\langle V_{SS}\rangle 
\end{array} \right) \left( \begin{array}{c}
K_{2A} \\ K_{2B} \end{array} \right) ,$$
where $M_0$ stands for the sum of the constituent quark masses in either case. 
The $V_{LS}^{-}$ term acts only on the $I=1/2$ singlet and triplet states 
giving rise to the spin-orbit mixing between these states\footnote{The 
spin-orbit $^3D_2-^1D_2$ mixing is a property of the model we are considering;
the possibility that another mechanism contributes to this mixing, such as
mixing via common decay channels \cite{Lip} should not be ruled out, but is not
included here.}, and is responsible for the physical masses of the $K_2$ and 
$K_2^{'}.$ Let us assume, for simplicity, that $$\sqrt{2}\langle V_{LS}^{-}
\rangle (K_{2B})\simeq -\sqrt{2}\langle V_{LS}^{-}\rangle (K_{2A})\equiv 
\Delta .$$ The masses of the $K_{2A},\;K_{2B}$ are then determined by 
relations similar to those for the $\pi _2,\;\rho _2$ above, and $M(K_2^{'})
\simeq M(K_{2A})+\Delta ,$ $M(K_2)\simeq M(K_{2B})-\Delta ,$ or\footnote{
Actually, as follows from Eq. (28) below, $$\frac{M(K_2^{'})-M(K_{2A})}{M(K_{
2B})-M(K_2)}=\frac{M(K_2)+M(K_{2B})}{M(K_2^{'})+M(K_{2A})}\simeq \frac{2M(K_{
2B})}{2M(K_{2A})}\simeq 1,$$ when both the deviations $M(K_{2B})-M(K_2),$ $M(
K_2^{'})-M(K_{2A})$ and the mass difference $M(K_{2A})-M(K_{2B})$ are small 
compared to $M(K_{2A}),\;M(K_{2B}).$}
\beq
\Delta \simeq M(K_2^{'})-M(K_{2A})\simeq M(K_{2B})-M(K_2).
\eeq
We thus obtain the following formulas for the masses of all eight $I=
1,1/2$ $D$-wave mesons, $\pi _2,\rho ,\rho _1,\rho _2,K_{2B},K^\ast ,K_{2A},K_
3^\ast :$
\bqry
M(^1D_2) & = & m_1+m_2-\frac{3}{4}\frac{a}{m_1m_2}, \\
M(^3D_1) & = & m_1+m_2+\frac{1}{4}\frac{a}{m_1m_2}-\frac{3b}{m_1m_2}-\frac{c}{2
m_1m_2}, \\
M(^3D_2) & = & m_1+m_2+\frac{1}{4}\frac{a}{m_1m_2}-\frac{b}{m_1m_2}+\frac{c}{2
m_1m_2}, \\
M(^3D_3) & = & m_1+m_2+\frac{1}{4}\frac{a}{m_1m_2}+\frac{2b}{m_1m_2}-\frac{c}{7
m_1m_2},
\eqry 
where $a,b$ and $c$ are related to the matrix elements of $V_{SS},$ $V_{LS}$
and $V_T$ (see Eqs. (4), (6), (10)) and assumed to be the same for all of 
the $D$-wave states, and we have ignored the correction to $V_{LS}^{+}$ in the
formula (10) that is due to the difference in the masses of the $n$ and $s$ 
quarks. These masses, as calculated from (12)-(15), are
(in the following, $\pi _2$ stands for the mass of the $\pi _2,$ etc., and we
assume $SU(2)$ flavor symmetry, $n\equiv m_u=m_d,$ $s\equiv m_s)$
\beq
n=\frac{5\pi _2+3\rho +5\rho _2+7\rho _3}{40},
\eeq
\beq
s=\frac{10K_{2A}+6K^\ast +10K_{2B}+14K_3^\ast -5\pi _2-3\rho -5\rho _2-7\rho _
3}{40}.
\eeq
With the physical values of the meson masses (in GeV), $\pi _2\cong 1.67,$ 
$\rho \simeq \rho _2\simeq \rho _3\cong 1.70,$ $K_{2A}\simeq K_{2B}\cong 1.80,$
$K^\ast \simeq K_3^\ast \cong 1.77,$ the above relations give $$n\simeq 850\;
{\rm MeV,}\;\;\;s\simeq 940\;{\rm MeV,}$$ so that the abovementioned 
correction, according to (10), is $\sim 90^2/(2\cdot 850\cdot 940)\simeq 
0.5$\%, i.e., completely negligible. It follows from (12)-(15) that
\bqry
\frac{15a}{m_1m_2} & = & 3M(^3D_1)+5M(^3D_2)+7M(^3D_3)-15M(^1D_2), \\ 
\frac{60b}{m_1m_2} & = & 14M(^3D_3)-5M(^3D_2)-9M(^3D_1), \\ 
\frac{30c}{7m_1m_2} & = & 5M(^3D_2)-2M(^3D_3)-3M(^3D_1). 
\eqry
By expressing the ratio $n/s$ in four different ways, viz., directly from 
(16),(17) and dividing the expressions (18)-(20) for the $I=1/2$ and $I=1$ 
mesons by each other, one obtains the three relations,
$$\frac{5\pi _2+3\rho +5\rho _2+7\rho _3}{10K_{2A}+6K^\ast +10K_{2B}+14K_3^
\ast -5\pi _2-3\rho -5\rho _2-7\rho _3}$$
\beq
=\frac{3K^\ast +5K_{2A}+7K_3^\ast -15K_{2B}}{3\rho +5\rho _2+7\rho _3-15\pi _
2},
\eeq
\beq
\frac{3K^\ast +5K_{2A}+7K_3^\ast -15K_{2B}}{3\rho +5\rho _2+7\rho _3-15\pi _
2}=\frac{14K_3^\ast -5K_{2A}-9K^\ast }{14\rho _3-5\rho _2-9\rho },
\eeq
\beq
\frac{14K_3^\ast -5K_{2A}-9K^\ast }{14\rho _3-5\rho _2-9\rho }=\frac{5K_{2A}-
2K_3^\ast -3K^\ast }{5\rho _2-2\rho _3-3\rho }.
\eeq
First consider Eq. (23) which may algebraically be rewritten as
\beq
(K_3^\ast -K^\ast )(\rho _3-\rho _2)=(K_3^\ast -K_{2A})(\rho _3-\rho ).
\eeq
Since the $\rho $ and $\rho _3$ states are mass near-degenerate, $\rho \approx
\rho _3$ (their masses are $1700\pm 20$ MeV and $1691\pm 5$ MeV, respectively
\cite{pdg}), it then follows from (24) that either $\rho _2\approx \rho 
\approx \rho _3,$ or $K^\ast \approx K_3^\ast .$ The first possibility leads, 
through the relations (19),(20) applied to the $I=1$ mesons, to $b\approx c
\approx 0,$ which would in turn, from the same relations for the $I=1/2$ 
mesons, imply $K^\ast \approx K_{2A}\approx K_3^\ast .$ Although this case may 
not be excluded on the basis of current experimental data on the meson masses,
we consider simultaneous disappearance of both the spin-orbit and tensor terms
as dubious. We believe, therefore, that the physical case corresponds to
\beq
K^\ast \approx K_3^\ast ,
\eeq
so that, the mass near-degeneracy of the 1 $^3D_1$ and 1 $^3D_3$ meson nonets 
in the $I=1$ channel, $\rho \approx \rho _3,$ implies similar near-degeneracy 
also in the $I=1/2$ channel. This result is a direct consequence of the model 
we are considering; the equality $K^\ast =K_3^\ast $ follows from Eq. (24), 
independent of the values of the input parameters $a,b,c,n,s,$ with the 
proviso that the result $\rho =\rho _3$ is borne out experimentally. 

With $K^\ast =K_3^\ast $ and $\rho =\rho _3,$ Eqs. (21) and (22) may be 
rewritten as
\beq
(\rho -\rho _2+K^\ast -K_{2A})(\pi _2+\rho _2+2\rho )=2(K^\ast -K_{2A})(
K_{2A}+K_{2B}+2K^\ast ),
\eeq
\beq
(K_{2A}-K_{2B})(\rho -\rho _2)=(K^\ast - K_{2A})(\rho _2-\pi _2).
\eeq  
One now has to determine the values of $\rho _2,$ $K_{2A}$ and $K_{2B}.$ The 
remaining equation is obtained from the mixing of the $K_{2A}$ and $K_{2B}$
states which results in the physical $K_2$ and $K_2^{'}$ mesons. Independent 
of the mixing angle, 
\beq
K_{2A}^2+K_{2B}^2=K_2^2+K_2^{'2}.
\eeq
With (in MeV) $\pi _2=1670\pm 20,$ $\rho=\rho _3\cong 1690,$ $K^\ast =K_3^\ast
\cong 1780,$ $K_2=1773,$ $K_2^{'}=1816,$ the solution to (26)-(28) is
\beq
\rho _2=1741\mp 19\;{\rm MeV,}\;\;\;K_{2A}=1827\mp 17\;{\rm MeV,}\;\;\;
K_{2B}=1762\pm 18\;{\rm MeV.}
\eeq
For this solution, we observe the sum rule
\beq
K_{2A}^2-\rho _2^2=0.307\;{\rm GeV}^2\simeq K_{2B}^2-\pi _2^2=0.316\;{\rm GeV
}^2,
\eeq
which may be further generalized to include the near-degenerate $\rho \approx
\rho _3\cong 1690$ MeV and $K^\ast \approx K_3^\ast \cong 1780$ MeV:
\beq
K^{\ast 2}-\rho ^2\approx K_3^{\ast 2}-\rho _3^2\cong 0.312\;{\rm GeV}^2.
\eeq  
Relations of the type (30),(31) could have been expected by anology with the
formulas $$K^{\ast 2}-\rho ^2=K^2-\pi ^2,\;\;\;K_2^{\ast 2}-a_2^2=K^2-\pi ^2,
\;\;\;{\rm etc.,}$$ provided by either the algebraic approach to QCD \cite{OT}
or phenomenological formulas $$m_1^2=2Bn+C,\;\;\;m_{1/2}^2=B(n+s)+C$$ (where
$B$ is related to the quark condensate, and $C$ is a constant within a given 
meson nonet) motivated by the linear mass spectrum of a nonet and the 
collinearity of Regge trajectories of the corresponding $I=1$ and $I=1/2$
states, as discussed in ref. \cite{linear}. 

Note from (29) that both the $K_{2A}$ and $K_{2B}$ lie in the mass 
intervals provided by current experimental data on the $K_2^{'}$ and $K_2$ 
states, respectively. This simply means that the mixing between these states is
negligible (within uncertainties provided by data), or $\sqrt{2}\langle V_{LS}^
{-}\rangle <<K_{2A}-K_{2B}.$ As we will see in Eqs. (32)-(34) below, this is
entirely consistent with reasonable expectation based on the decrease of such
matrix elements with increasing partial wave (see the corresponding $P$-wave
results \cite{prev}).
  
Thus, the nonrelativistic constituent quark model we are considering suggests
the following $q\bar{q}$ assignments for the isovector and isodoublet states 
of the $D$-wave meson nonets:
\bqryn
\pi _2 & \simeq  & 1680\;{\rm MeV,}\;\;\;K_{2B}\;\simeq \;1770\;{\rm MeV,} \\
\rho  & \simeq  & 1690\;{\rm MeV,}\;\;\;K^\ast \;\;\simeq \;\;1780\;{\rm MeV,}
 \\
\rho _2 & \simeq  & 1730\;{\rm MeV,}\;\;\;K_{2A}\;\simeq \;1820\;{\rm MeV,} \\
\rho _3 & \simeq  & 1690\;{\rm MeV,}\;\;\;K^\ast \;\;\simeq \;\;1780\;{\rm 
MeV.}
\eqryn
Let us now extract the matrix elements of the spin-spin, 
spin-orbit, and tensor interaction in our model. As follows from (18)-(20) and
the above relations for the masses of the $I=1,1/2$ mesons, 
\bqry
\langle V_{SS}\rangle  & \simeq  & \frac{a}{n^2}\;\simeq \;\frac{a}{ns}\;\cong
\;23.3\;{\rm MeV}, \\
\langle V_{LS}^{+}\rangle  & \simeq & \frac{b}{n^2}\;\simeq \;\frac{b}{ns}\;
\cong -3.3\;{\rm MeV}, \\
\langle V_T\rangle  & \simeq & \frac{c}{n^2}\;\simeq \;\frac{c}{ns}\;\cong \;
46.7\;{\rm MeV}.
\eqry
Also, $\langle V_{LS}^{-}\rangle \cong 0,$ since the $K_{2A}-K_{2B}$ mixing 
angle is close to zero. Therefore, the spin-spin  and tensor terms of the
Hamiltonian (2) are of the same order of magnitude, and the spin-orbit terms 
are negligibly small.  

One may now estimate the masses of the isoscalar mesons of the four nonets
assuming that they are pure $s\bar{s}$ states. Applying (12)-(15) with $m_1=m_
2=s,$ we find
\beq
\eta _2\simeq 1860\;{\rm MeV},\;\;\;\phi \approx \phi _3\simeq 1870\;{\rm MeV},
\;\;\;\phi _2\simeq 1910\;{\rm MeV}.
\eeq
The value 1870 is within 1\% of the physical value of the $\phi _3$ mass,
$1854\pm 7$ MeV \cite{pdg}. There exists an experimental candidate for the 
$\eta _2(1860)$ but it was omitted from the recent Meson Summary Table as 
``needs confirmation''. This state indicated in PDG as the $\eta _2(1870)$ 
\cite{pdg} has been seen by the Crystal Ball collaboration in the final state 
$\eta \pi ^0\pi ^0$ of a $\gamma \gamma $ reaction as a resonant structure 
having mass and width $1881\pm 32\pm 40$ MeV, $221\pm 92\pm 44$ MeV, 
respectively \cite{Karch}, and as a similar structure in $\gamma \gamma 
\rightarrow \eta \pi ^{+}\pi ^{-}$ by the CELLO collaboration, with mass and 
width $1850\pm 50$ MeV, $\sim 360$ MeV, respectively \cite{Feindt}.      

The masses of the remaining isoscalar $n\bar{n}$ states of the four nonets may 
be calculated by assuming that all four nonets are ideally mixed and using the
Sakurai mass formula for an ideally mixed nonet \cite{Sak},
\beq
M^2(I=1)+M^2(I=0,n\bar{n})+2M^2(I=0,s\bar{s})=4M^2(I=1/2).
\eeq
In this way, one obtains
\beq
\eta _2^{'}\simeq 1670\;{\rm MeV,}\;\;\;\omega \approx \omega _3\simeq 1680\;
{\rm MeV,}\;\;\;\omega _2\simeq 1720\;{\rm MeV.}
\eeq
The value 1680 is within 1\% of the physical value of the $\omega _3$ mass,
 $1667\pm 4$ MeV, and 2\% of that of the $\omega ,$ $1649\pm 24$ MeV \cite{
pdg}. 

\section{Concluding remarks}
We have shown that a nonrelativistic constituent quark model displays a common
mass near-degeneracy of the 1 $^3D_1$ and 1 $^3D_3$ meson nonets in the 
isovector and isodoublet channels, and suggests therefore that the $K^\ast (
1680)$ cannot be the $I=1/2$ member of the 1 $^3D_1$ nonet. The mass of the 
true member of the latter is estimated to be $\simeq 1780$ MeV. This may 
support the assumption of T\"{o}rnqvist that the $K^\ast (1680)$ should resolve
into two separate resonances which are the $I=1/2$ members of the 1 $^3D_1$ 
and 2 $^3S_1$ nonets. The analysis of the LASS data on the reaction $K^{-}p
\rightarrow \bar{K}^0\pi ^{-}p$ done by Bird \cite{Bird} reveals a resonant
structure with mass $1678\pm 64$ MeV and a huge width of $454\pm 270$ MeV; the
two abovementioned states may be associated with its upper- and lower-mass
parts, respectively. 

The conclusion that the $K^\ast (1410)$ does not belong to the 2 $^3S_1$ 
nonet agrees with the results obtained by one of the authors in ref. \cite{LB} 
on the basis of the linear spectrum of a meson nonet discussed in \cite{
linear}, which does not support the $K^\ast (1410)$ meson being the member of 
the 2 $^3S_1$ nonet. (In \cite{LB}, out of the two, $K^\ast (1410)$ and $K^\ast
(1680),$ the preference being the 2 $^3S_1$ $I=1/2$ state was given to the
latter). If this is actually the case, and the true member of the 2 $^3S_1$
nonet is, e.g., the low-mass part of the broad $K^\ast (1680),$ in agreement 
with T\"{o}rnqvist, the question immediately arises as to what the real nature
of this state is, if it does exist. A possible answer to this question may be
the subject of subsequent investigation. \\
     
We close with briefly summarizing our findings: \\
1. A nonrelativistic constituent quark model displays a common mass 
near-degeneracy of the 1 $^3D_1$ and 1 $^3D_3$ meson nonets in the $I=1$ and 
$1/2$ channels, and suggests therefore that the $K^\ast (1680)$ cannot be the
$I=1/2$ member of the 1 $^3D_1$ nonet. \\
2. When matched to current experimental data on the meson masses, this model 
shows no mixing between the $I=1/2$ states of the 1 $^3D_2$ and $^1D_2$ nonets.
The spin-orbit terms of the Hamiltonian appear to be negligibly small. \\
3. The results suggest a sum rule $$M^2(I=1/2)-M^2(I=1)\approx \;{\rm const}
\simeq 0.31\;{\rm GeV}^2,$$ which holds for all four $D$-wave meson nonets. \\
4. The results also suggest that the $\eta _2(1870)$ which is at present 
omitted from the Meson Summary Table, is the $I=1$ $s\bar{s}$ state of the 1
$^1D_2$ nonet. \\
5. The $q\bar{q}$ assignments for the $P$-wave nonets obtained on the basis
of the results of the work, are 
 
1 $^1D_2$ $J^{PC}=2^{-+},$ $\;\pi _2(1680),$ $\eta _2^{'}(1670),$ $\eta _2(
1860),$ $K_{2B}(1770)$

1 $^3D_1$ $J^{PC}=1^{--},$ $\;\rho (1690),$ $\;\;\omega (1680),$ $\;\phi (
1870),$ $\;K^\ast (1780)$ 

1 $^3D_2$ $J^{PC}=2^{--},$ $\;\rho _2(1730),$ $\omega _2(1720),$ $\phi _2(
1910),$ $K_{2A}(1820)$

1 $^3D_3$ $J^{PC}=3^{--},$ $\;\rho _3(1690),$ $\omega _3(1680),$ $\phi _3(
1870),$ $\;K_3^\ast (1780)$

\section*{Acknowledgments}
Correspondence of one of the authors (L.B.) with L.P. Horwitz during the 
preparation of this work is greatly acknowledged.

\end{document}